\documentclass[twocolumn,floatfix,showpacs,preprintnumbers,amsmath,amssymb,prb]{revtex4}
\usepackage[dvips]{graphicx}\usepackage{psfrag}
\usepackage{dcolumn}
\usepackage{bm}
\usepackage{amsbsy}

\newcommand{\be}{\begin{equation}}
\newcommand{\ee}{\end{equation}}
\newcommand{\bea}{\begin{eqnarray}}  
\newcommand{\eea}{\end{eqnarray}}

\newcommand{\p}{\partial}
\newcommand{\s}{\sigma}

\newcommand{\la}{\langle}
\newcommand{\ra}{\rangle}
\newcommand{\rd}{\mbox{d}}
\newcommand{\ri}{\mbox{i}}

\newcommand{\eps}{\epsilon}
\newcommand{\nn}{\nonumber}
\newcommand{\bp}{\bar{\partial}}
\newcommand{\bz}{\bar{z}}

\newcommand{\bxi}{\bar{\xi}}

\begin{document}

\input{epsf}
\title{Ordered phases of XXZ-symmetric spin-1/2 zigzag ladder}
\author{M. Zarea$^1$, M. Fabrizio$^1$, and A.A. Nersesyan$^{2,3}$}
\affiliation{
{$^1$ International School for Advanced Studies, Via Beirut 4, 34014, Trieste, Italy}\\
{$^2$ The Abdus Salam International Centre for Theoretical Physics,
P.O.Box 586, 34100, Trieste, Italy}\\
{$^3$ The Andronikashvili Institute of Physics, Tamarashvili 6, 380077, Tbilisi,
Georgia}
}
\date{\today}
\begin{abstract}
Using bosonization approach, we derive an effective low-energy theory for
XXZ-symmetric
 spin-1/2 zigzag ladders and discuss its phase diagram by a variational
approach. A spin nematic phase emerges in a wide part of the phase diagram,
either critical or massive. Possible crossovers between the spontaneously
dimerized and spin nematic phases are discussed, and the topological
excitations in all phases identified.
\end{abstract}

\pacs{75.10.Pq, 75.40.-s, 75.30.Gw, 71.10.Pm}
\maketitle 
\section{Introduction}
\medskip
The appearance of unconventional spin-liquid phases 
in frustrated Heisenberg models around 
a critical point separating 
different magnetic ordered phases is 
a long standing,
intriguing issue which has attracted  
notable theoretical and experimental interest in recent years\cite{sachdev}. 
From a theoretical point of view, this is quite a challenging problem which 
calls for a deep reexamination of 
standard theories. Conventionally, the stability of 
a spin-ordered phase may be investigated by spin-wave theory or by more 
sophisticated field-theoretical approaches based on the non-linear $\sigma$-model. 
Spin-wave theory is in principle able to detect instabilities at any wave-vector, although 
a reliable description would require a systematic $1/S$ expansion, $S$ being the 
magnitude of the spin. The non-linear $\sigma$-model 
offers a better description of the 
critical behavior close to an instability point, yet it has a major limitation. Namely, 
it takes into account only the long-wavelength Goldstone modes 
of the ordered state under consideration, but fails to describe the excitations 
at the wave-vector of the competing ordered state. Hence it does not provide any information 
about the phase which can emerge under increasing the effect of frustration.

One-dimensional spin models might turn useful to attempt an improvement of the 
field-theoretical approach in view of an extension to higher dimensions.
The simplest example of one-dimensional frustrated spin model  
is the spin-1/2 Heisenberg chain with  antiferromagnetic nearest neighbor 
exchange $J\,'$ and
frustrating next nearest neighbor exchange $J$ (for a recent review and 
references therein, see Ref.\onlinecite{leche-review}).
Besides the general interest, this
model is also relevant for realistic materials, 
such as $Cs_2CuCl_4$,
in which magnetic $Cu$ ions get arranged into a zig-zag fashion.\cite{coldea2}

In the classical limit, the $J$-$J\, '$ spin chain has a Ne\`el long range order 
for $j=J/J\,' < 1/4$, with characteristic momentum $q=\pi/a_0$. 
In the Ne\`el phase time reversal symmetry is preserved if 
combined with a translation by one lattice spacing $a_0$.
For larger values of $j$, a spiral ordering at momentum satisfying
$\cos(q\, a_0)=-1/4\,j$ is stabilized at the classical level. 
There parity and time reversal symmetries are separately broken.
 
Quantum fluctuations modify the classical phase 
diagram \cite{haldane}. By Mermin-Wagner's theorem, 
spin rotational symmetry cannot be broken in one dimension: the Ne\`el 
long-range ordered phase turns into a quasi-long range ordered one
characterized by 
power-law decaying 
correlations of the staggered magnetization. 
The low-energy 
effective critical theory is the level-1 Wess-Zumino-Novikov-Witten (WZNW) 
model (free massless bosons with central charge $C=1$). 
On the contrary, the spiral order disappears completely 
in favor of a 
spontaneoulsy dimerized phase. The transition point is sligthly 
shifted
with respect to the classical value, $j_c \simeq 0.241$ {\cite{Eggert,Okamoto}}. 
Within the WZNW 
model 
formalism, the transition 
is driven by 
a perturbation which is marginally irrelevant (relevant) at
$j<j_c$ ($j>j_c$). At $j=j_c$ 
a Berezinskii-Kosterlitz-Thouless transition takes place, 
and an exponentially small spectral gap opens up in the region $j>j_c$\cite{affleck}. 
The system continuously passes to a two-fold 
degenerate, spontaneously dimerized,  massive phase. Upon
further increasing $j$, the gap 
reaches its maximum 
at the exactly solvable Majumdar-Ghosh point \cite{Ghosh}, $j=1/2$, after which it 
slowly decreases \cite{affleck}.
Even though the ground state of the $J$-$J'$ chain remains dimerized at all
$j > j_c$, above the Majumdar-Ghosh point the system reveals signatures
of the classical spiral phase: 
the spin-spin correlations become incommensurate, as it was shown numerically 
in Refs.\onlinecite{affleck,essler}.
Since this occurs
far away from the 
region of applicability of the SU(2)$_1$ WZNW 
model, there is little 
scope to improve the field-theoretical description of the gapless phase at $j<j_c$ to account 
for incommensurate correlations 
that emerge
well above $j_c$. More promising is to 
approach this problem from the opposite side, $j\gg 1$. If $J\, '=0$, the even and odd 
sublattices 
of the spin chain
decouple, and the model effectively describes two decoupled Heisenberg chains, 
one for each sublattice. 
Classically this corresponds to the case when the spiral 
wave number is equal to $q=\pi/2a_0$. 
Switching on a small $J\, '$ transforms the model to a weakly coupled
two-chain zigzag spin ladder,
with the interchain coupling giving rise to a 
marginally relevant perturbation that
opens up a gap and brings the system back to the dimerized phase. 
In addition it should also move the 
relevant 
momentum $q$ away from $\pi/2a_0$ towards $\pi/a_0$. Therefore incommensuration and 
the spectral gap are supposed to 
appear together in this limit, which makes a field-theoretical 
description more plausible. 

Indeed, in the limit $j\gg 1$, a novel, parity-breaking (twist) perturbation
was identified 
in Ref.~\onlinecite{nge} as a natural sourse of the spin incommensurabilities.
The twist term has a tendency to support a finite spin current along the chains
which would account for the expected shift of the momentum. 
However, for the SU(2)-symmetric zigzag ladder, the situation still remains rather unclear.
Apparently, the appearance
of a nonzero spin current is not compatible with
the requirement of unbroken spin rotational symmetry
(see, however, the discussion in sec.VII).
On the other hand, no reliable information about the actual role of the twist
operator at the strong-coupling fixed point can be extracted from the 
Renormalization Group (RG) analysis \cite{nge} because of the perturbative nature of
this approach.
Thus, the structure of the low-energy effective
field theory for the SU(2)-symmetric S=1/2 zigzag ladder still remains unknown.

The situation changes much to the better in the presence of
strong spin anisotropy. Namely,
close to the XX limit, 
a self-consistent, symmetry-preserving mean-field approach shows that the twist operator 
can stabilize a new, spin-nematic (chiral)  phase \cite{nge}.
In this doubly degenerate phase  
a nonzero spin current polarized along the 
easy axis flows along the ladder, and the transverse spin-spin correlations are
incommensurate and may even decay algebraically within some parameter range, as 
it is the case at the
XX point.
This picture is supported by recent numerical simulations \cite{kabu,kabu1,pedro1,pedro2}.
In particular,  the numerical work by
Hikihara {\em et al} \cite{kabu,kabu1} 
has indeed confirmed the existence of the 
critical spin-nematic phase in a broad region of the phase diagram for  
spin-anisotropic chains, both for integer and half-integer spins.
NMR experiments with compound $Ca V_2 O_4$ \cite{fuku} having a spin-1 double-chain zigzag
magnetic structure have revealed the gapless nature of the spectrum, which may be
an indication to the critical chiral state of this material.
On the other hand, 
numerical simulations 
have revealed the existence of 
a new gapped chiral phase in a very narrow region between the dimerized phase and 
the critical chiral phase, but, within numerical accuracy, only for integer spins. 
In that gapped phase, the spin  current coexists with dimerization; accordingly, the
spin-spin correlations are incommensurate but decay 
exponentially.
Recently the phase diagram for general spin $S$ has been 
studied by 
bosonization technique {\cite{leche}}  and by means of the non-linear $\sigma$-model 
{\cite{kole}}, verifying the existence of both critical and gapped chiral 
phased for integer spin.

In this paper we present a detailed study of the phase diagram 
of the XXZ-symmetric, frustrated, spin-1/2 
chain in the limit $J'\ll J$. 
Using a variational analysis of the bosonized Hamiltonian we identify 
possible phases of the model. In addition to the critical spin-nematic 
phase and to the commensurate spontaneously dimerized one, we find 
conditions for the existence of a massive spin-nematic region 
for the $S=1/2$ case. We also characterize the 
topological excitations which occur in each region of the phase diagram.   

In the following section we introduce the model and discuss the bosonization 
approach. In section III we demonstrate how the variational approach 
can be applied to the twistless ladder, in which only the dimerization operator
plays a role.  Critical spin nematic phase, driven only by the twist operator,
is studied
in section IV. The interplay between dimerization and twist operator and
other emerging phases is 
discussed in 
section V. In section VI we discuss a ferromagnetic 
phase which turns out to be  
dual to the critical
spin nematic phase. In section VII the RG approach is implemented to study
the interplay between different twist operators at the border of these
mutually dual phases.
The last section contains conclusions.

\section{The model and its low-energy limit}
We consider a frustrated spin-1/2 Heisenberg chain with $2L$ sites, described by the 
Hamiltonian
\be
H = \sum_{a=x,y,z}\sum_{n=1}^{2L} \left[ J'_a \, S^a _n\, S^a _{n+1} 
+ J_a\, S^a _n\, S^a _{n+2} \right],
\label{Ham:singlechain}
\ee
where
\bea
J_x = J_y = J > 0, && J_z = J \Delta, \nonumber\\
J' _x = J' _y = J' > 0, &&
J' _z =  J' \Delta'.
\label{couplings}
\eea
In what follows, $\Delta$ and $\Delta'$ 
will be treated as independent anisotropy parameters.
Upon the transformation
\[
{\bf S}_{2n} \to {\bf S}_1(n),\;\;\;
{\bf S}_{2n+1} \to {\bf S}_2(n),
\]
the model (\ref{Ham:singlechain}) is mapped onto the zig-zag spin-1/2 ladder 
Hamiltonian 
\bea
H &=& \sum_{a=x,y,z} \sum_{n=1}^L \sum_{i=1,2}
J_a S^a _i (n) S^a _i (n+1) ~
\nonumber\\
 &+& \sum_{a=x,y,z} \sum_n J' _a \left[S^a _1 (n) + S^a _1 (n +1) \right]
S^a _2 (n) 
\label{ham}
\eea

Let us discuss some general properties of this model.
In the limit $J=0$, (\ref{Ham:singlechain}) describes a standard Heisenberg 
antiferromagnetic chain where the spin-spin correlations are modulated with 
wavevector $q=\pi$. On the contrary, when $J'=0$, the equivalent model (\ref{ham}) describes two 
decoupled 
spin chains. The modulating wavevector in this case is $q=\pi/2$. When both 
$J$ and $J'$ are finite, we may expect two possible behaviors of the spin structure factor 
$S(q)=(2L)^{-1}\sum_{n,m} \langle {\bf S}_{n} \cdot {\bf S}_{m}\rangle {\rm e}^{iq(n-m)}$: 
either it is peaked at $q=\pi$ and $q=\pi/2$, or it shows a single peak at an incommensurate 
$q_0$ which smoothly moves from $q=\pi/2$ at $J\gg J'$ towards $q=\pi$ when $J'\gg J$. 
Translated into the zig-zag ladder language, the former case implies that, for $n$ large,  
\bea 
&&\langle {\bf S}_1(n) \cdot {\bf S}_1(0) \rangle = 
\langle {\bf S}_2 (n)\cdot {\bf S}_2(0) \rangle \nonumber\\
&&~~~~~~ = F_0(|2n|) + F_\pi(|2n|) 
+ (-1)^n\, F_{\pi/2}(|n|),
\nonumber\\
&&\langle {\bf S}_1(n) \cdot {\bf S}_2(0) \rangle = F'_0(|2n+1|) -F'_\pi(|2n+1|), 
\nonumber\\
&&\langle {\bf S}_2(n) \cdot {\bf S}_1(0)\rangle = F'_0(|2n-1|) -F'_\pi(|2n-1|), 
\nonumber
 \eea
where $F_0(|n|)$, $F_\pi(|n|)$ and $F_{\pi/2}(|n|)$, as well as the primed ones, 
are smooth real functions describing the contributions of the 
$q=0$, $q=\pi$ and $q=\pi/2$ modes, respectively. 
The difference between 
the 1-2 and 2-1 spin-spin 
correlators, as well as 
the 
absence of inversion symmetry 
$n\to -n$, reflect the fact that 
the zigzag ladder lacks
two $Z_2$ symmetries -- the $1 \leftrightarrow 2$ interchange symmetry and
site parity $P_S$ (understood as $P^{(1)}_S \otimes P^{(2)}_S$). 
However, if the model is gapless or possesses a small spectral gap inducing a
macroscopically large correlation length,
then 
site-parity is effectively 
restored at long distances.

If, apart from $q=0$, the spin structure factor has a peak 
at an incommensurate wave vector $q_0 \in [\pi/2,\pi]$, then we expect that 
\bea 
&&\langle {\bf S}_1 (n) \cdot {\bf S}_1(0) \rangle = 
\langle {\bf S}_2(n)\cdot {\bf S}_2(0) \rangle \nonumber \\
&&~~~~~~~~~~~~ = F_0(|2n|) + 
F_{q_0}(|2n|) \, \cos 2nq_0,
\nonumber\\
&&\langle {\bf S}_1(n) \cdot {\bf S}_2(n) \rangle = F'_0(|2n+1|) \nonumber \\
&& ~~~~~~~~~~~~+ F'_{q_0}(|2n+1|)\,\cos q_0 (2n+1),\label{1-2} \\ 
&&\langle {\bf S}_2(0) \cdot {\bf S}_1(n) \rangle = F'_0(|2n-1|) \nonumber \\
&& ~~~~~~~~~~~~+ F'_{q_0}(|2n-1|)\,\cos q_0 (2n-1).\label{2-1}
\eea
We notice that the presence of the modulating factors in (\ref{1-2}),(\ref{2-1})
makes the the breakdown of $P_S$ even more pronounced and, contrary to the
commensurate case, this breakdown will
survive the continuum limit we are 
going to adopt. 
Thus, the two different types of spin correlations -- commensurate or
incommensurate -- can be distinguished within a continuum, low-energy description
by an asymptotic restoration or breakdown of the site-parity symmetry.

As discussed in the Introduction, in this paper we are going to study the model 
(\ref{ham}), or equivalently (\ref{Ham:singlechain}), in the limit 
$J\gg J'$ of weakly coupled chains. That allows us to adopt
the well-known continuum description of each XXZ chain based on the
bosonization approach \cite{lp} (see also Ref. \onlinecite{gnt} for a recent review) and 
then treat the interchain coupling 
as a weak perturbation. 
Bosonization of the XXZ zigzag spin-1/2 ladder has already been discussed
in Ref.\onlinecite{cabra}. Here we review this procedure in more detail paying attention
to the structure of the effective continuum model
which is important for the subsequent analysis of the phase diagram.

\medskip

We start with the Abelian bosonization of a single XXZ spin-1/2 chain.
Its universal, low-energy properties in the gapless 
Luttinger liquid phase ($-1 < \Delta \leq 1$) are adequately described by
the Gaussian model for a massless scalar field $\varphi(x) = \varphi_R (x) 
+ \varphi_L (x)$, 
\bea
&& H_{\rm XXZ} \to  \int \rd x~ {\cal H}_G (x), \nn\\ 
&& {\cal H}_G = (v_s /2) 
\left[Q^{-1} \left(\p_x \varphi   \right)^2  + Q
\left(\p_x \vartheta  \right)^2  \right] 
\label{xxz-cont}
\eea
Here 
$\vartheta (x) = - \varphi_R (x) + \varphi_L (x) $ 
is the field dual to $\varphi(x)$,
$\pi(x) = - \p_x \vartheta (x)$ being the momentum conjugate to $\varphi(x)$, 
$v_s \sim J a_0$ is the velocity of the collective spin excitations,
and $Q$ is the the Luttinger liquid parameter 
which determines the compactification
radius of the field $\varphi$. The dependence $Q = Q(\Delta)$ 
in the whole
range $-1 < \Delta \leq 1$ is known from the Bethe-ansatz solution: 
$1/Q = 1 - (1 / \pi) \arccos \Delta$.
Thus $Q$ varies in the range $\infty > Q \geq 1$ when $\Delta$ takes values
within the interval $-1 < \Delta \leq 1$. In particular, $Q=2$  
at the XX point $(\Delta = 0)$, and $Q=1$ at the SU(2)-symmetric (Heisenberg) point
$(\Delta = 1)$. 
Throughout this paper it will be assumed that
$\Delta < 1$ (i.e. $Q > 1$). In this case the perturbation to the Gaussian model 
(\ref{xxz-cont}),
$
\lambda_U \cos \sqrt{8\pi} \varphi
~~(\lambda_{U} \sim J\Delta),
$
which in terms of the Jordan-Wigner fermions originates from Umklapp processes,
is 
strongly 
irrelevant and will be dropped in what follows.

Since for each chain only the uniform and staggered low-energy modes
survive the continuum limit, the corresponding spin densities 
can be parametrized as follows:
\bea
{\bf S}_{i}(n) &\to& a_0 {\bf S}_i(x),~~ (x=na_0) \nn\\
{\bf S}_i(x) &=& {\bf J}_{i,R} (x) + {\bf J}_{i,L} (x)
+ (-1)^n {\bf n}_i (x)~~(i=1,2). \nn
\eea
Here $a_0$ is the lattice spacing, ${\bf J}_{i,R,L}$ 
are chiral components of the smooth part of the magnetization
of the $i$-th chain, and
${\bf n}_i$ is the staggered magnetization. The latter is even under
site parity transformation ($P_S$) and odd under link parity transformation
($P_L$). A distinctive feature of the zigzag ladder 
is that it is invariant
under mixed parity: 
$P^{(1)}_S \otimes P^{(2)}_L$ and
$P^{(1)}_L \otimes P^{(2)}_S$. By this symmetry,  strongly relevant terms, $n^a _1  n^a _2$,
which determine the spin-liquid properties
of the unfrustrated spin-1/2 ladders \cite{snt}, are instead forbidden in model 
(\ref{ham}). As a result, in the low-energy limit, the interchain perturbation,
$
{\cal H}' = {\cal H}_{JJ} + {\cal H}_{\rm twist},
$
is contributed by the ``current-current'' interaction\cite{affleck,allen-senechal}
\be
{\cal H}_{JJ} = 2 \sum_{a=x,y,z} g_a J^a _1 J^a _2, \label{cur-cur}
\ee
and also by ``twist'' terms\cite{nge} allowed by the $P^{(1)}_S \otimes P^{(2)}_L$
symmetry:
\be
{\cal H}_{\rm twist} 
= \frac{1}{2} \sum_{a=x,y,z} g_a a_0 T^a + 
\frac{1}{2} g_0 a_0 T^0, \label{twist}
\ee
Here
\bea
T^a = n^a _1 \p_x n^a _2 - n^a _2 \p_x n^a _1, ~~~~
T^0 = \eps_1 \p_x \eps_2 - \eps_2 \p_x \eps_1,\label{T-scal}
\eea
are chirally asymmetric operators with conformal spin 1, and 
$\eps_i \simeq (-1)^n {\bf S}_{i}(n) \cdot {\bf S}_{i}(n+1)$ represent the continuum limit 
of the dimerization operators. (Note that for a single chain 
$\eps (x)$ is even under $P_L$ and odd under $P_S$.) 
The coupling constants are given by
$
g_x = g_y \equiv g_{\perp} = J' a_0,
$ 
$
g_z \equiv g_{\parallel} = J' \Delta' a_0.
$

The ``current-current'' and twist perturbations are of different nature.
The former are parity 
[i.e. $P^{(1)}_{S(L)} P^{(2)}_{S(L)}$]
symmetric. If acting alone, provided that the interchain exchange
is antiferromagnetic, these lead to spontaneous dimerization of the ground state
(see section III), the existence of massive
topological excitations (spinons), and the onset of short-ranged {\sl commensurate} 
interchain spin correlations \cite{affleck,aen}.
The twist terms,
whose appearance stems from the frustrated nature of interchain interaction, 
explicitly break parity. However, by the previous discussion, parity can be broken 
either in a mild way, which 
is the case when the leading asymptotics of spin-spin correlations are still commensurate,
or
more profoundly, i.e. explicitly inducing 
{\sl incommensurations} in the spin correlations. Both patterns of the low-energy behavior
of the system will be discussed below.

In model (\ref{ham}), 
only the vector part of the twist perturbation, $g_a T^a$,
emerges in the continuum limit. The scalar part, $g_0 T^0$, although absent 
in the bare Hamiltonian, 
is generated in the course of RG flow\cite{nge}.

For this reason we will assume that such a term is present at the outset, with a bare 
amplitude $g_0$.

Let us first bosonize ${\cal H}_{JJ}$. In terms of the rescaled fields,
$\phi_i = (1/\sqrt{Q})\varphi_i$ and $\theta_i = {\sqrt{Q}}\vartheta_i$,
the 
``currents'' $J^a _i = J^a _{i,R} + J^a _{i,L}$, are given 
by \cite{com1}
\be
J^z _{i} 
= \sqrt{\frac{Q}{2\pi}} \p_x \phi_{i}, ~~
J^+ _{i} = - \frac{\zeta}{\pi\alpha} e^{i \sqrt{\frac{2\pi}{Q}} \theta_i}
\cos \sqrt{2\pi Q} \phi_i ,
\label{J}
\ee
where $\alpha$ is the short-distance cutoff of the bosonic theory, 
and $\zeta(Q)$ is a nonuniversal (and yet unknown) positive 
constant approaching the value 1 in the SU(2) limit.
Using the definitions (\ref{J}) and 
passing to the symmetric and antisymmetric combinations of the fields,
$\phi_{\pm} = (\phi_1 \pm \phi_2)/\sqrt{2}$,
$\theta_{\pm} =(\theta_1 \pm \theta_2)/\sqrt{2}$,
we find that
the longitudinal ($zz$) part of ${\cal H}_{JJ}$ 
adds to the Gaussian part of the model transforming the latter into
\be
{\cal H}_G \to \sum_{\s = \pm} \frac{v_{\s}}{2} \left[
R_{\s} (\p_x \theta_{\s})^2  + R_{\s}^{-1} (\p_x \phi_{\s})^2  
\right], \label{quadra}
\ee
with
\be
\frac{1}{R_{\pm}} = \frac{v_{\pm}}{v_s} = \sqrt{1 \pm \frac{g_{\parallel}Q}{\pi v_s}}
= 1 \pm \frac{g_{\parallel}Q}{2\pi v_s} + O(g^2 _{\parallel}).
\label{param-quadr} 
\ee
The exact dependence of $R_{\pm}$
on the dimensionless parameter $g_{\parallel}Q/\pi v_s$ is unknown.
Therefore we will restrict ourselves to the case
$|g_{\parallel}|Q /\pi v_s \ll 1$ and keep only linear terms in the expansion
(\ref{param-quadr}). For a weak interchain interaction ($|g_{\parallel}|/\pi v_s \ll 1$),
this is justified almost for the whole range $|\Delta| < 1$
except for a narrow region $\Delta + 1 \sim (g_{\parallel}/\pi v_s)^2$
close to the ferromagnetic transition point. 
The parameters $R_{\pm}$ then satisfy the relation
$
R \equiv R_+ = 1/R_- 
$
which considerably simplifies the perturbative analysis.

Performing an additional rescaling of the fields,
$
\phi_{\pm} = \sqrt{R_{\pm}} \Phi_{\pm}, 
$
$
\theta_{\pm} = \sqrt{R_{\mp}} \Theta_{\pm},
$
for the transverse (xx, yy) part of ${\cal H}_{JJ}$
one finds:
\bea
&&{\cal H}_{JJ; \perp} = 2 g_{\perp} \sum_{a=x,y} J^a _1 J^a _2
= \frac{\lambda_{\perp}}{\pi\alpha}
\left( {\cal D} + {\cal F} \right), \label{perp-part}\\
&& {\cal D} = \cos \sqrt{4\pi K_+} \Phi_+ \cos \sqrt{4\pi K_-} \Theta_- , \nn\\
&& {\cal F} = \cos \sqrt{4\pi/ K_-} \Phi_-
\cos \sqrt{4\pi K_-} \Theta_- ,
\label{D/F-op}
\eea
where $\lambda_{\perp} = g_{\perp}\zeta^2 /\pi\alpha$, and 
\be
K_+ = QR, ~~~K_- = R/Q.
\label{K-pm}
\ee

To bosonize the twist perturbation (\ref{twist}),
we use the bosonization formulas
for the staggered magnetization  of the S=1/2 XXZ chain (see e.g. \onlinecite{gnt}):
\bea
n^z _i &=& - \left( C_z / \pi \alpha\right) \sin \sqrt{2\pi Q}~ \phi_i \nn\\
n^{\pm} _i &=& \left( C_x / \pi \alpha\right)
\exp (\pm \ri \sqrt{2\pi/Q}~ \theta_i), 
\label{n-bosoniz}
\eea
where $C_a (Q) ~(a=x,z)$
are noniniversal parameters (their exact dependence on $Q$ was
recently found in Refs.~\onlinecite{lz,l}). 
Then, in terms of the fields $\Phi_{\pm}$, $\Theta_{\pm}$,
the twist term becomes:
\be
{\cal H}_{\rm twist} = \sum_{i=1,2,3} \lambda_i {\cal O}_i,
\label{twist-boson}
\ee
with
\be
\lambda_1 \sim C^2 _x g_{\perp}/\alpha, ~~~ 
\lambda_{2,3} \sim C^2 _z (\pm g_{\parallel} + g_0)/\alpha,
\label{twist-const}
\ee
 and three bosonized twist operators ${\cal O}_{1,2,3}$ related to
$T^i ~(i=0,1,2,3)$ as follows:
\bea
&&{\cal O}_1 = T^x + T^y = \frac{2}{\sqrt{K_+}}\p_x \Theta_+ 
\sin \sqrt{4\pi K_-}\Theta_- ,
\label{tw-1}\\
&&{\cal O}_2 = \frac{T^0 + T^z}{2} = \sqrt{K_+}
\p_x \Phi_+ \sin \sqrt{\frac{4\pi} {K_-}} \Phi_- , \label{tw-2}\\
&&{\cal O}_3 = \frac{T^0 - T^z }{2}
= \frac{1}{\sqrt{K_-}} \p_x \Phi_- \sin \sqrt{4\pi K_+} \Phi_+ \label{tw-3}.
\eea

 Thus, the bosonized continuum version of our model, 
${\cal H} = {\cal H}_0 + {\cal H}_{JJ;\perp} + {\cal H}_{\rm twist}$, 
represents a Gaussian field theory of two scalar fields, 
\be
{\cal H}_0 = \sum_{\s=\pm} {\cal H}^{(\s)}_0
= \sum_{\s=\pm} \frac{v_{\s}}{2}
\left[ \left( \p_x \Phi_{\s} \right)^2
+   \left( \p_x \Theta_{\s} \right)^2 \right],
\label{Gaussian-models}
\ee
with perturbations (\ref{perp-part}) and (\ref{twist-boson}) which couple the $(+)$ and $(-)$
channels together. Since ${\cal H}_0$ is perturbed in a relevant way,
the relationship between the coupling
constants of the original model (\ref{Ham:singlechain}) and the parameters of ${\cal H}$,
obtained within our weak-coupling approach, is not to be trusted.
For this reason we will 
consider those parameters as independent. Namely, we will 
treat the Hamiltonian ${\cal H}$ as a 
low-energy effective theory for a most general class of
frustrated zigzag spin-1/2 ladders, sharing the same
symmetry properties with the model (\ref{Ham:singlechain}).
\medskip

The scaling dimensions of the perturbing operators are:
\bea
&& 
d_{\cal D} = K_+ + K_- , ~~
 d_{\cal F} = K_- + \frac{1}{K_-} , \nn\\
&& d_1 = 1 + K_-, ~~
 d_2 = 1 + \frac{1}{K_-},~~
d_3 = 1 + K_+ . \label{dims}
\eea
Their relevance ($d < 2$) or irrelevance ($d > 2$)
can be understood
from Fig.1 where the plane ($K_-, K_+$) is shown.
The point $K_+ = K_- = 1$ corresponds to the 
SU(2)-symmetric zig-zag ladder where
all perturbations (including the Umklapp term) 
are marginal\cite{nge}. 
This point and its close vicinity will not be discussed in this paper.
Due to the condition $K_+/K_- = Q^2 \geq 1$,
the physical part of the ($K_-, K_+$) plane 
lies above the line $K_+ = K_-$ and can be divided into four sectors in which
at least one twist operator is relevant:

\bea
&&{\rm sector~A}:~~~ 
d_1 < 2, ~~~~ d_{\cal F} \geq 2,~~d_2, d_3, d_{\cal D} > 2,   \nn\\
&&{\rm sector~B}:~~~ 
d_1 <  d_{\cal D} < 2, ~~~~~~~~~d_2, d_3, d_{\cal F} > 2,  \nn\\
&&{\rm sector~ C:}~~~
d_{\cal D} < d_1 <  d_3 < 2, ~~~~~~~d_2, d_{\cal F} > 2, \nn\\
&&{\rm sector~ D:}~~~
d_2 < 2, ~~~~ d_{\cal F} \geq 2,~~d_1, d_3, d_{\cal D} > 2 . \label{sectors}
\eea

\begin{figure}[h]
\centering
\psfrag{K1}{$K_+$}\psfrag{K2}{$K_-$}\psfrag{Di}{${\cal D}$}
\psfrag{O1}{${\cal O}_1$}\psfrag{O2}{${\cal O}_2$}\psfrag{O3}{${\cal O}_3$}
\psfrag{Massive Spin Nematic}{MSN}
\psfrag{A}{A}\psfrag{B}{B}\psfrag{C}{C}\psfrag{D}{D}\psfrag{1}{1}\psfrag{0}{0}\psfrag{a}{}\psfrag{b}{}
\includegraphics[height=7cm]{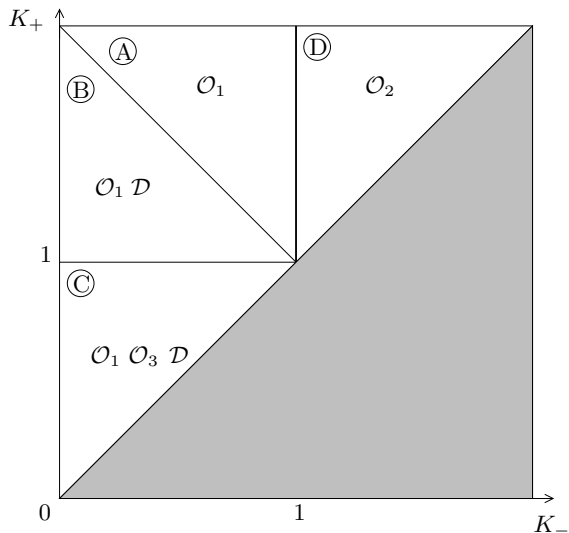}
\caption{The parameter space of the model}
\end{figure}

Notice that, except for the operator ${\cal D}$, all other perturbing operators 
have a nonzero
Lorentz spin: $S_{1,2,3} = 1$, $S_{\cal F} = 2$. Strictly speaking, 
the conventional criterium of relevance does not apply to such operators (see e.g. 
Ref.~\onlinecite{gnt}) because in higher orders of perturbation theory they
can generate
relevant scalar perturbations.
Using standard fusion rules for the 
Gaussian model, we have analyzed the structure of various terms appearing in
the second order of perturbation theory. There are marginal terms leading to  small
corrections to the parameters $K_{\pm}$ and the velocities $v_{\pm}$, as well as those
which renormalize the already existing coupling constants.
Besides, new scalar operators $\cos 2 \sqrt{4\pi K_{+}} \Phi_+$,
$\cos 2\sqrt{4\pi/K_-} \Phi_-$, $\cos 3 \sqrt{4\pi K_-} \Theta_-$
and
$\cos 2 \sqrt{4\pi K_-} \Theta_-$ are generated.
The first two of
them have scaling dimensions $4QR$ and $4Q/R$, respectively, and are therefore strongly
irrelevant (since $Q>1$, $R \sim 1$). The third operator has dimension
$9 K_- = 9 R/Q$ and becomes relevant roughly at $Q > 9/2$, which is a region
of the ferromagnetic intrachain exchange, far away from the XX point.
This region will not be considered here. Finally, the last 
perturbation $ \sim \cos 2 \sqrt{4\pi K_-} \Theta_-$
has dimension
$4K_- = 4R/Q$ and thus becomes relevant at $K_- < 1/2$, which
corresponds to a vicinity of the XX point $Q = 2$. 
However, even in that case its role is subdominant, as we will show later.
\medskip

Thus, the continuum model we will be dealing with in the remainder of
this paper reads:
\be
{\cal H} = {\cal H}_0 - (\lambda_{\perp}/\pi\alpha) {\cal D} 
+ 
\sum_{i=1,2,3} \lambda_i {\cal O}_i.
\label{cont-model-final}
\ee
where ${\cal H}_0$ is given by (\ref{Gaussian-models}).
For later purposes, we have suitably inverted the sign of the coupling
constant $\lambda_{\perp}$ by making a shift of the field $\Phi_+$:
$
\Phi_+ \to \Phi_+ + \sqrt{\pi/4K_+}. \label{shift+}
$
In what follows, we will analyze possible phases of this model
in the four sectors A,B,C,D by using a generalization of the standard variational 
approach \cite{coleman} that 
accounts for ground states with nonzero values
of topological charges, $\la \p_x \Theta_- \ra$ and $\la \p_x \Phi_{\pm} \ra$.
The very possibility to incorporate such states within the variational method
stems from the fact that the twist operators (\ref{tw-1})--(\ref{tw-3})
are products of 
fields belonging to different Gaussian models ${\cal H}^{\pm}_0$.

\section{Twistless ladder}
Before addressing the role of the twist terms in (\ref{cont-model-final})
it is instructive 
first to apply the variational approach to a simpler
frustrated two-leg ladder model\cite{twistless,aen}
which, in the continuum limit,
is free from parity-breaking perturbations yet 
being spontaneously dimerized.
This model is the two leg-ladder version of the standard $J_1$-$J_2$ frustrated 
Heisenberg plane, in which the interchain coupling includes besides the usual
on-rung coupling, $J_{\perp}$, a frustrating exchange, $J_{\times}$, across the diagonals
of the plaquettes. 
In the XXZ case its Hamiltonian reads:
\bea
H_{\rm gen} &=& 
\sum_{a,n} \sum_{i=1,2}
J_a  S^a _i (n) S^a _i (n +1) 
+ \sum_{a,n} J' _a S^a _1 (n) S^a _2 (n) \nn\\  
&+& \sum_{a,n}  J'' _a \left[ S^a _1 (n) S^a _2 (n+1) +  S^a _1 (n+1) S^a _2 (n) \right],
\label{twistless}
\eea
where $J_a$ are defined as in (\ref{couplings}), and
\bea
J' _x = J' _y = J_{\perp}, ~&&~ J' _z = J_{\perp}\Delta', \nn\\
J'' _x = J'' _y = J_{\times}, ~&&~ J'' _z = J_{\times}\Delta' .\nn
\eea
The $P^{(1)}_{S(L)}\otimes P^{(2)}_{S(L)}$ reflection symmetry of the model (\ref{twistless})
forbids the marginal twist perturbations to appear in the continuum limit.
The additional condition $J_{\perp} = 2 J_{\times}$ eliminates the 
$n^a _1 n^a _2$ part of the interchain coupling and thus makes the two decoupled Gaussian
models 
only perturbed by the current-current interchain
interaction (\ref{cur-cur}) with coupling constants
$g_{\perp} = 2 J_{\times}\Delta' a_0$ and $g_{\parallel} = 2 J_{\times} a_0$.
So the bosonized continuum Hamiltonian has the structure of
 Eq.(\ref{cont-model-final}) with
$\lambda_{\perp} \neq 0$ and $\lambda_{1,2,3} = 0$.
As we shall see below, the dimerized phase it describes also occurs in the most
part of sector C of the zigzag ladder, Eq.(\ref{cont-model-final}), where
${\cal D}$ is the most relevant perturbation to the Gaussian models
(\ref{Gaussian-models}).

\medskip

To implement Coleman's variational procedure\cite{coleman}, we 
introduce a trial ground state,
\be
| {\rm vac}\ra = |0; m_+, m_- \ra = |0; m_+ \ra \otimes |0; m_- \ra ,
\ee
which describes free  bosons 
in the $(\pm)$ sectors with masses $m_{\pm}$. These are regarded 
as variational parameters. To estimate the variational
ground-state energy density,
$
E_0 (m_+, m_-) = \la {\rm vac}| {\cal H} | {\rm vac} \ra,
$
one needs to normal order the Hamiltonian  
with the prescription that, in the normal-mode expansions 
of $\Phi_{\pm}(x)$ and $\Theta_{\pm}(x)$,
$m_{\pm}$ should be treated as infrared regulator masses.

Upon normal ordering\cite{coleman} 
\bea
&&{\cal H}^{(\pm)}_0 = {\cal N}_{m_{\pm}} \left[ {\cal H}^{(\pm)}_0 \right]\nn\\
&& + \frac{1}{8\pi v_{\pm}}
\left[ 2 \left(\frac{v_{\pm}}{\alpha}\right)^2 +   m^2 _{\pm} + 
O \left(\frac{m^2 _{\pm} \alpha^2}{v^2 _{\pm}}\right) \right], 
\label{reordering1} \\
&& 
\cos \sqrt{4\pi K_+} \Phi_+
= \left( \frac{|m_{+}|\alpha}{v_{+}}\right)^{K_+}
{\cal N}_{m_{+}}\left[ \cos \sqrt{4\pi K_+} \Phi_+ \right],
\nn\\
&&
\cos \sqrt{4\pi K_-} \Theta_-
= \left( \frac{|m_{-}|\alpha}{v_{-}}\right)^{K_-}
{\cal N}_{m_{-}}\left[ \cos \sqrt{4\pi K_-} \Theta_- \right],
\nn
\eea
where ${\cal N}_{m_{\pm}}$ are the normal ordering symbols.
Subtracting from (\ref{reordering1}) the diverging contribution of the
zero-point motion when $\alpha \to 0$, 
ignoring for simplicity the difference between the velocities $v_{\pm}$ 
and defining the dimensionless quantities,
\be
{\cal E} = \frac{4\pi \alpha^2}{v} E_0, ~~M_{\pm} = \frac{m_{\pm}\alpha}{v},
~~z_{\perp} = \frac{4\lambda_{\perp}\alpha}{v},
\label{notations}
\ee
we find that
\be
{\cal E} (M_+, M_-) = \frac{M^2 _+ + M^2 _-}{2} - z_{\perp} |M_+|^{K_+}|M_-|^{K_-}.
\label{var-en-twistless}
\ee
Withous loss of generality we choose $M_{\pm}$ to be positive.
With the condition $z_{\perp} \ll 1$ in mind, it should be understood
that the masses  $M_{\pm}$, obtained upon minimization of ${\cal E}$,
should satisfy $M_{\pm} \ll 1$.
\medskip

At $d_{\cal D} > 2$ only a trivial solution exists, $M_{\pm} = 0$,
corresponding to a critical regime in which the interchain interaction
is irrelevant and the two chains asymptotically decouple in the low-energy
limit. At $d_{\cal D}=K_+ + K_- <2$ we find a nontrivial solution,
\be
\frac{M_+}{\sqrt{K_+}} = \frac{M_-}{\sqrt{K_-}}
= K_+ ^{\frac{K_+}{2(2-d_{\cal D})}}K_- ^{\frac{K_-}{2(2-d_{\cal D})}}
z_{\perp}^{\frac{1}{2-d_{\cal D}}},
\label{masses-twistless}
\ee
with ground state energy given by:
\bea
{\cal E}_D &=& - \left(  \frac{1 - K_+}{2K_+}\right) M^2 _+ -
\left(  \frac{1 - K_-}{2K_-}\right) M^2 _- 
\label{D.phase-energy}\\
&=& - \left(  \frac{2 - d_D}{2K_+}\right) M^2 _+ \label{D.phase-energy1}
\eea
This solution describes a strong-coupling, massive phase 
in which the fields $\Phi_{+}$ and $\Theta_-$
are locked in one of infinitely degenerate minima of the potential 
${\cal U}(\Phi_+,\Theta_-) = - (\lambda_{\perp}/\pi\alpha) {\cal D}$.
Since $\lambda_{\perp} > 0$, these minima decouple into ``even'' and ``odd''
sets:
\bea
\Phi_{+} = \sqrt{\frac{\pi}{4K_+}}2n_+, ~~~~~~~~ \Theta_- = \sqrt{\frac{\pi}{4K_-}}2n_-;
~~~~~~~\nn\\
\Phi_+ = \sqrt{\frac{\pi}{4K_+}}\left( 2n_+ + 1\right), ~
\Theta_- = \sqrt{\frac{\pi}{4K_-}}\left( 2n_- + 1\right),
\eea
where $n_{\pm} = 0,\pm1,\pm2, ...$. 
The existence of these two inequivalent sets reflects two-fold degeneracy
of the spontaneously dimerized ground state. Transverse dimerization is the
order parameter; it is defined as
$\la \eps_{\perp} (x) \ra$ where \cite{affleck,aen} 
\bea
&&\eps_{\perp}(x) = {\bf n}_1(x) \cdot {\bf n}_2(x)  
\propto C^2 _x \cos \sqrt{4\pi K_-} \Theta_- 
\nn\\
&&+ \frac{1}{2} C^2 _z \left( \cos \sqrt{4\pi K_+} \Phi_+
+ \cos \sqrt{4\pi/K_-} \Phi_- \right).
\label{dim-o.p.}
\eea
Since the field $\Theta_-$ is locked, its dual $\Phi_-$ is disordered
and, hence, the expectation value of the last term in (\ref{dim-o.p.}) vanishes.
Hence, 
\be
\la \eps_{\perp} \ra = \pm \eps_0, ~~
\eps_0 \propto C^2 _x |M_-|^{K_-} + \frac{1}{2} C^2 _z |M_+|^{K_+},
\label{eps-fin}
\ee
with the two signs of $\eps$ corresponding to the even and odd vacua, 
respectively.
\medskip

The discrete ($Z_2$) symmetry that is spontaneously broken 
in the ground state is generated by even-odd interset transitions of the
fields, 
\be
\Delta \Phi_{+} = \pm \sqrt{\pi/4 K_{+}},~~
\Delta \Theta_- = \pm \sqrt{\pi/4 K_{-}},
\label{interset-tr}
\ee
and is related to translations by one lattice
spacing on one chain only. This is not an exact symmetry of the microscopic Hamiltonian
(\ref{twistless}) but rather appears as an important property of the 
corresponding low-energy model with a ``current-current'' perturbation.
The excitation spectrum of the model consists of pairs of massive
topological kinks (spinons) interpolating between two adjacent minima of the
potential ${\cal U}$. The kinks carry two topological quantum numbers -- 
the total spin 
\be
S^z _{+} = 
\sqrt{\frac{K_+}{\pi}} \int_{-\infty}^{\infty} \rd x~ \p_x \Phi_+ (x),
\ee
and the relative longitudinal spin current
\be
j^z _{-}/u =  - \sqrt{\frac{K_-}{\pi}}\int_{-\infty}^{\infty} \rd x~ \p_x \Theta_- (x),
\label{rel-curr}
\ee
which, according to (\ref{interset-tr}), take fractional values 
$\pm 1/2$ \cite{comment}.


\section{Critical spin nematic phase}



Now we are coming back to the continuum 
model (\ref{cont-model-final}) for the XXZ zig-zag ladder.
We begin our discussion with sector A where
the twist operator ${\cal O}_1$ is the only relevant perturbation
to the two decoupled Gaussian models ${\cal H}^{(\pm)}_0$. 
Making a shift
$
\Theta_- \to \Theta_- +  (1/4) \sqrt{\pi/K_-}
$
we write the low-energy model in sector A as follows: 
\be
{\cal H}_A  = {\cal H}^{(+)}_0 + {\cal H}^{(-)}_0
+ \frac{2\lambda_1}{\sqrt{K_+}} \p_x \Theta_+ \cos\sqrt{4\pi K_-} \Theta_-.
\label{ham-sn} 
\ee
This model has the same structure as that for the XX zigzag ladder considered
in Ref.\onlinecite{nge}. Not surprisingly, the variational procedure we will follow now
leads to qualitatively the same results as those obtained for the XX case
within a symmetry-preserving mean-field approach\cite{nge}. 
\medskip

The interaction term in (\ref{ham-sn}) couples the vertex operator
in the (--) channel to the topological current density $\p_x \Theta_+$ in
the (+) channel. The latter determines 
the z-component of the spin current which flows along the chain direction, $\hat{j}^z_{||}$, 
\be
\hat{j}^z _{+}\equiv \hat{j}^z_{||}= - v \sqrt{\frac{K_+}{\pi}} \p_x \Theta_+.
\label{long-spin-current}
\ee 
We observe that a finite $\lambda_1$, see Eq. (\ref{ham-sn}), generates an additional 
contribution to the spin current, $\hat{j}^z_\perp$, which flows along the interchain bonds.
By the continuity equation related to the conservation of the $z$-component of the total spin, 
one finds that 
\be
\hat{j}^z_\perp = - \frac{2}{\sqrt{\pi}}\, \lambda_1 \, \cos\sqrt{4\pi K_-} \Theta_-.
\label{def:jperp}
\ee
The total spin current is therefore $\hat{j}^z = \hat{j}^z_{||} + \hat{j}^z_\perp$, and  
the twist operator ${\cal O}_1$ is nothing but a coupling 
term $\hat{j}^z_{||}\, \hat{j}^z_\perp$.

The structure of the perturbation in the model (\ref{ham-sn})
suggests that the ground state admits finite values of the
mass gap  in the (--) channel
{\sl and} the spin current in the (+) channel.
So one needs to treat both of these two quantities
as variational parameters. 
To this end, we keep boundary conditions periodic for the field $\Theta_- (x)$
but impose twisted boundary conditions for the field $\Theta_+ (x)$:
\be
\Theta_+ (x) = \Theta_+ ^0 (x) - \frac{1}{v}\sqrt{\frac{\pi}{K_+}}{j}^z _{+} x.
\label{theta-curr}
\ee
Here $\Theta_+ ^0 (x)$ is a massless harmonic Bose field satisfying
periodic boundary conditions, and  ${j}^z _{+}$ is the average value
of the current operator (\ref{long-spin-current}) which is to be determined
self-consistently.
The variational procedure is the same as in the previous section
with the exception that the ground state energy in the (+) channel will
acquire a piece proportional to $({j}^z _{+})^2$.
Otherwise this sector remains gapless: $M_+ = 0$.
Using dimensionless notations, 
\be
{\cal J}_{+} = \frac{2\pi \alpha}{\sqrt{K_+} v} {j}^z _{+},
 ~~~z_1 = 4 \sqrt{\frac{\pi}{K_+}} \frac{\lambda_1 \alpha}{v},
\label{dim-less-sn}
\ee
for the variational energy density ${\cal E}$ we obtain:
\be
{\cal E} ({\cal J}_{+}, M_- )
= \frac{1}{2} \left(M^2 _- + {\cal J}^2 _{+}\right)
\mp z_1 {\cal J}_{+} M^{K_-} _{-}.
\label{var-en-sn}
\ee
As before, we have chosen $M_-$ to be positive.  The $(\mp)$ signs in the
interaction term correspond to two sets of vacuum expectation values
of the field $\Theta_-$: $\Theta_- = \sqrt{\pi/K_-} n$ and 
$\Theta_- = \sqrt{\pi/K_-} (n+1/2)$, respectively.

\medskip

Minimizing ${\cal E}$ with respect to $M_{\pm}$ and ${\cal J}_{+}$
we find that the (--) channel is gapped, 
\be
M_- = K_{-}^{\frac{1}{2(1-K_-)}}z_1 ^{\frac{1}{1-K_-}}, \label{mass-sn}
\ee
if $K_- < 1$. This is actually the condition $d_1 < 2$ for the twist operator
${\cal O}_1$ to be a relevant perturbation, which is satisfied 
in sector A. At the same time, the gap supports a finite
value of the spin current in the (+) channel:
\be
{\cal J}_{+} = \pm \frac{M_-}{\sqrt{K_-}}
=
\pm K_- ^{\frac{K_-}{2(1-K_-)}}z_1 ^{\frac{1}{1-K_-}}.
\label{curr-sn}
\ee

We notice that the dimensionless transverse current defined by 
\be
\hat{\cal J}_{\perp} = - \frac{2\pi \alpha}{\sqrt{K_+} u}\hat{j}^z _{\perp}
= z_1 \cos \sqrt{4\pi K_-} \Theta_- ,\label{dim-less-perp-curr}
\ee
also acquires a finite ground-state expectation value 
\be
{\cal J}_{\perp} = \la \hat{\cal J}_{\perp} \ra
= \mp z_1 M_- ^{K_-}, \label{ave-trans-curr}
\ee
which exactly cancels ${\cal J}_{||}={\cal J}_{+}$, so that the total spin current is zero. 
This results in a spin nematic 
(or a staggered spin-flux) phase
characterized by local spin currents circulating around elementary plaquettes
in an alternating way. This type of ordering does not break time reversal
symmetry.
In sector A the spin nematic phase is critical because the spin-density fluctuations in the (+)
channel remain gapless.

We notice that the transverse current can be associated with the chirality order parameter.
The latter is defined as
\be
\kappa_z = \la \kappa_z (x) \ra,
~~~
\kappa_z  (x) = \left[ {\bf n}_1 (x) \times {\bf n}_2 (x) \right]_z ,
\label{chirality-o.p.}
\ee
and, according to bosonization rules (\ref{n-bosoniz}),
transforms in the continuum limit to
\be
\kappa_z  (x) \propto \cos \sqrt{4\pi K_-} \Theta_- (x) \propto \hat{j}_{\perp}(x). 
\label{kappa-bos}
\ee

As the  dimerized phase discissed in sec.III,
the spin nematic phase is doubly degenerate because the mixed parity symmetry
$P^{(1)}_{S(L)} P^{(2)}_{L(S)}$ is spontaneously broken in the ground state.
The two degenerate phases differ in the signs of the longitudinal and transverse
currents.
Consequently, apart from the massless bosonic mode describing low-energy
fluctuations of the total magnetization,
there exist massive topological $Z_2$ kinks 
corresponding to vacuum-vacuum transitions,
\[
{\cal J}_{+} \to - {\cal J}_{+}, ~~~
\Theta_- \to \Theta_- \pm \sqrt{\pi/4 K_-},
\]
and thus carrying the
relative spin current
${j}^z _{-}/u = \pm 1/2$; see Eq.(\ref{rel-curr}).

\medskip

The presence of a finite longitudinal spin current in the ground state
makes the transverse (xy) spin correlations incommensurate. Since the (+) channel is
massless and described by a Gaussian field with a $K_+$ dependent compactification
radius, the correlations will decay algebraically with a nonuniversal
exponent. Making use of bosonization rules (\ref{n-bosoniz}),
Eq. (\ref{theta-curr}) and the fact
that the field $\Theta_-$ is locked, one easily finds the asymptotic behaviour of 
the transverse  spin correlation function:
\be
\la S^+ _{1} (x) S^- _{1,2} \ra \propto \frac{(-1)^{x/a_0}}{|x|^{1/2K_+}} e^{-iq_0 x}
\label{incom-corr}
\ee
where the wave vector $q_0 = \pi {\cal J}_{\parallel}/u K_+$.
At the XX point ($Q=2$, $R=1$, $K_+ = 2$) the spin correlations decay according to
the power law $|x|^{1/2}$, in agreement with Ref. \onlinecite{nge}.
\medskip

The ground state energy of the critical spin nematic (CSN) phase is given by
\be
{\cal E}_{\rm CSN} = - \left( \frac{1-K_-}{2K_-} \right) M^2 _-
\label{crit-SN-energy}
\ee

Before closing this section, we would like to briefly discuss the role of the scalar
operator $\cos 2\sqrt{4\pi K_-}\Theta_-$, which is generated in a higher orders and
becomes relevant at  $K_- < 1/2$. In the presence of a finite spin current in (+) 
channel, this term transforms the effective Hamiltonian in (--) channel 
to a double-frequency sine-Gordon model:
 ${\cal H}_0 - \lambda_{\rm eff}\sin\sqrt{4\pi K_-} \Theta_-
 - g\cos 2 \sqrt{4\pi K_-} \Theta_-$,
where $\lambda_{\rm eff} \propto \lambda_1 \la j^z _+ \ra$. 
It is known \cite{dm,fgn} that
the $g$-term can induce an Ising transition to a new massive phase
if $g>0$ and $g^{1/2(1-2K_-)} > \lambda_{\rm eff}^{1/(2-K_-)}$.
The last inequality, however, is not satisfied since the amplitude $g \sim \lambda^2 _{\rm eff}$ 
is rather small and, hence, the presence of the second harmonics does not qualitatively affect the
above results.


\section{Massive spin nematic and dimerized phases}


Let us now move to sectors B and C where the properties of the systems
are determined by the interplay between two most relevant perturbations,
${\cal O}_1$ and ${\cal D}$. The second twist perturbation,
${\cal O}_3$, is either irrelevant (as in sector B) or the least relevant 
(as in sector C). In Appendix A we explicitly show that 
its role is indeed subdominant
in sectors B and C far from 
the SU(2)-symmetric point, $K_+ = K_- = 1$, 
\medskip

Thus, the effective Hamiltonian reads:
\bea
{\cal H}_{B/C} &=& {\cal H}^{(+)}_0 + {\cal H}^{(-)}_0\nn\\
&-& \frac{\lambda_{\perp}}{\pi\alpha} \cos\sqrt{4\pi K_+} \Phi_+ \cos\sqrt{4\pi K_-} \Theta_-
\nn\\
&+& \frac{2\lambda_1}{\sqrt{K_+}} \p_x \Theta_+ \sin\sqrt{4\pi K_-} \Theta_-
\label{ham-B}
\eea
The potential in (\ref{ham-B}) contains both the sine and cosine of the field $\Theta_-$;
so its vacuum value $\Theta^* _-$ is expected to be located somewhere within the interval
$(0, \sqrt{\pi/4K_-})$ and must be such that in a massive phase with $M_- \neq 0$
\bea
&&\la {\rm vac} | \cos \sqrt{4\pi K_-} \Theta^* _- | {\rm vac} \ra = M^{K_-}_-, \nn\\
&&\la {\rm vac} | \sin \sqrt{4\pi K_-} \Theta^* _- | {\rm vac} \ra = 0.
\eea
Setting $\Theta_- = \Theta_- ^* - \gamma/\sqrt{4\pi K_-}$, we arrive at the following 
expression of the dimensionless variational energy:
\bea
&&{\cal E}({\cal J}_{\parallel}, M_+, M_-, \gamma)
= \frac{1}{2} \left(M^2 _+ + M^2 _- +  {\cal J}^2 _{+} \right)\nn\\
&&~~~~\mp z_1 {\cal J}_{+} M^{K_-} _- \sin \gamma
- z_{\perp} M^{K_+} _+ M^{K_-} _- \cos \gamma.
\label{var-energyB}
\eea
Its minimization with respect to $M_{\pm}$, ${\cal J}_{+}$
and the angle $\gamma$ yields the following set of equations:
\bea
&& M^{K_-}_- ( z_{\perp} M^{K_+}_+ \sin \gamma \mp z_1 {\cal J}_{+} \cos \gamma )
= 0, \label{Bset-1}\\
&& {\cal J}_{+} \mp z_1 M^{K_-}_- \sin \gamma = 0,
\label{Bset-2}\\
&& M_+ ( 1 - z_{\perp} K_+ M^{K_+ - 2} _+ M^{K_-} _- \cos \gamma ) = 0,
\label{Bset-3}\\
&& M_- ( 1 - z_{\perp}  K_- M^{K_+}_+ M^{K_- - 2} _-  \cos \gamma \nn\\
&& ~~~~~~\mp z_1 K_- {\cal J}_{+} M^{K_- - 2} _- \sin \gamma )
= 0. \label{Bset-4}
\eea

There are two obvious solutions of these equations in which only
one of the two perturbing operators is effective. In these solutions
the angle $\gamma$ takes two values:  $0$ and $\pi/2$.
The corresponding phases are, respectively: (i) a fully gapped D phase already 
described in sec.III,
with zero current (${\cal J}_{+} = 0$) and nonzero 
masses $M_{\pm}$
given by Eq.(\ref{masses-twistless}), and
(ii) a CSN phase with nonzero ${\cal J}_{+}$ and $M_-$
given by Eqs. (\ref{curr-sn}) and (\ref{mass-sn}).
\medskip

Eqs. (\ref{Bset-1})-(\ref{Bset-4}) admit one more solution
where the combined effect of the two relevant perturbations leads to
an intermediate value of the mixing angle $\gamma$, 
\be
\cos\gamma = \frac{M_+}{M_-} \sqrt{ \frac{K_-}{K_+}}, 
\label{mix-angle:B}
\ee
and a finite mass gap in the (+) channel,
\be
M_+ = (z_{\perp} \sqrt{K_+}/z_1)^{\frac{1}{1-K_+}}.
\label{mass+:B}
\ee
This is a noncritical or massive spin nematic (MSN) phase
characterized by the coexistence of a reduced spin current ${\cal J}_+$
\be
{\cal J}_+ = \left[{\cal J}_+ \right]_{\rm CSN} \sin \gamma
\label{reduced-current}
\ee
and a nonzero dimerization
\be
\eps_{\perp} \propto \pm \left[  C^2 _x M_- ^{K_-} \cos \gamma
+ \frac{1}{2} C^2 _z M_+ ^{K_+} \right].
\ee

An important observation is that the minimal value of the variational energy
(\ref{var-energyB}) is still given by expression (\ref{D.phase-energy}).
Therefore, the energies of the MSN and CSN phases are related as
\[
{\cal E}_{\rm MSN} = {\cal E}_{\rm CSN} + \frac{K_+ - 1}{2K_+} M^2 _+ .
\]
So in sector B ($K_+ > 1$) the MSN phase is energetically 
less favorable than the CSN phase, and the ground state
should be chosen between CSN and D phases. Accordingly, in sector C ($K_+ < 1$)
the competing phases are MSN and D.
\medskip

Consider first sector B. Here we need to compare the ground state energies
of the D and CSN phases given by Eqs.(\ref{D.phase-energy1}) and 
(\ref{crit-SN-energy}). These are of the same order when the mass
gaps of the two phases,  Eqs. (\ref{masses-twistless}) and (\ref{mass-sn}),
become comparable. Notice that the coupling constants $z_1$ and $z_{\perp}$
are both  proportional to $g_{\perp}$ and, hence are of the same order of
magnitude; their ratio is
\be
z_{\perp}/z_1 = C \sqrt{K_+}, \label{ratio}
\ee 
where $C$ is a nonuniversal number. Therefore, the
condition 
\be
z^{\frac{2}{2-d_D}}_{\perp} \sim z^{\frac{2}{2-d_1}}_1
\label{mass-condition}
\ee
can be satisfied only in some vicinity of the line $K_+ = 1$ where
scaling dimensions of the operators ${\cal D}$ and ${\cal O}_1$
become equal. 
\medskip

As already mentioned, it is not possible to establish
a precise relationship between the parameters 
of the original, microscopic model (\ref{ham}) and the effective low-energy
theory (\ref{cont-model-final}). As a result, 
the parameter $C$ in (\ref{ratio}) is unknown. Therefore we are forced to
consider two cases, $z_1 > z_{\perp}$ and $z_1 < z_{\perp}$, on equal
footing and draw plausible scenarios for each of them, leaving the final
choice to future numerical work.
\medskip

Setting $K_+ = 1 + \delta$ with $|\delta| \ll 1$, we find that the condition
(\ref{mass-condition}) translates to the relation
\be
\delta = (1-K_-) \frac{\ln (z_{\perp}/z_1) }{ \ln(1/ z_1 )}.
\label{boundary}
\ee
This relation determines a line $\delta = \delta (K_-)$
which lies entirely in sector B ($\delta > 0$) and is located
very close to the line $K_+ = 1$ only if $z_1 < z_{\perp}$.
Under this condition the relation (\ref{boundary}) determines a phase
boundary between the CSN and D states. The transition is of first order,
associated with discontinuities of
the spin current and dimerization order parameters. 
It can be easily shown that in the case $z_1 < z_{\perp}$ the D phase,
occupying a narrow region close to the line $K_+ = 1$, extends over the
whole C phase.

\medskip

In the opposite case, $z_1 \geq z_{\perp}$, Eq.(\ref{boundary}) has no solution
for $\delta > 0$, implying that CSN is a stable ground state in the whole
sector B. Moving to sector C opens a possibility for the MSN phase.
If $1-K_+ \ll 1$,
the condition $z_1 \geq z_{\perp}$ admits a small nonzero mass $M_+$ given
by (\ref{mass+:B}). Thus is sector C the upper boundary for the MSN phase
is $K_+ = 1$. The lower boundary is found from the requirement
$\cos \gamma < 1$ (see Eq. (\ref{mix-angle:B})). Within the logarithmic
accuracy, this brings us again to Eq.(\ref{boundary}), this time  for $\delta (K_-) < 0$.
\medskip

Thus, if the ratio $z_1 / z_{\perp} > 1$, then the CSN  and D phases are
``sandwiched'' by the MSN phase occupying a narrow region in sector C
\be
1 - (1-K_-) \frac{\ln (z_1 / z_{\perp}) }{ \ln(1/ z_1 )}
< K_+ < 1 \label{msn-range}
\ee
attached to the line $K_+$ (see Fig. 2). In all this region 
${\cal E}_{\rm MSN} < {\cal E}_{\rm D}$. The transitions that occur on the upper and lower
boundaries of the MSN phase are continuous. When moving 
from sector B to sector C through the MSN phase, the mixing angle $\gamma$ varies 
from $\pi/2$ to 0.
Correspondingly, the current ${\cal J}_+$ decreases from its nominal value
$[{\cal J}_+]_{\rm CSN}$ and vanishes at the lower boundary,
whereas the transverse dimerization $\eps_{\perp}$ increases from zero at the upper boundary
and reaches its value $\eps_0$, Eq.(\ref{eps-fin}), in the pure D phase at 
the lower boundary (see Fig. 2). A possible way of driving the ladder to pass 
through these phases is shown in Fig. 3.

\begin{figure}[h]
\centering
\psfrag{K1}{$K_+$}\psfrag{K2}{$K_-$}\psfrag{Dimer Phase}{D}
\psfrag{Critical Spin Nematic}{CSN}
\psfrag{Massive Spin Nematic}{MSN}
\psfrag{A}{}\psfrag{1}{1}\psfrag{0}{0}\psfrag{a}{}\psfrag{b}{}
\includegraphics[width=8cm]{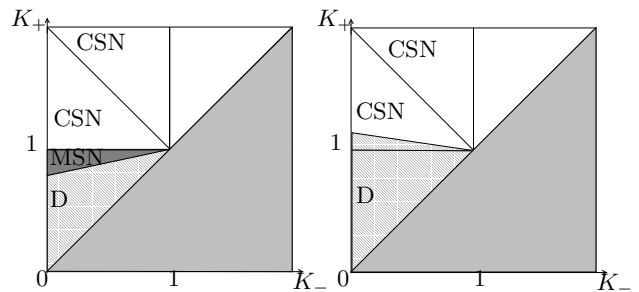}\label{newphase}
\caption{Possible phase transitions in the ladder: the left one is
 when $z_{\bot}<z_1$, MSN phase is very narrow, and the right one is for
  $z_{\bot}>z_1$ }
\end{figure}


\section{Ferromagnetic phase}
Let us now consider sector D where $K_+ > K_- > 1$. This condition implies that
$K_+ K_- = R^2 > 1$, and so this sector corresponds to the case of a
ferromagnetic interchain interaction ($g_{\parallel} < 0$).
The effective low-energy model
\be
{\cal H}_D = {\cal H}^{(+)}_0 + {\cal H}^{(-)}_0
+ \lambda_2 \sqrt{K_+} \p_x \Phi_+ \sin \sqrt{4\pi/K_-} \Phi_- \label{ham-f}
\ee
contains only one relevant twist operator and is dual to model (\ref{ham-sn})
describing the SN phase in sector A: 
mapping between these two models is achieved by duality transformations
\be
2\lambda_1 \to \lambda_2, ~~K_{\pm} \to 1/K_{\pm},~~\Phi_{\pm}
\to \Theta_{\pm} \label{dual-transf}
\ee
Using this correspondence, we can readily translate the results of sec.IV to
the present case. In particular, the spontaneously generated spin current
${\cal J}_+$ of the SN phase transforms to the z-component of the uniform
spin density. So the ground state of the system in sector D is
{\sl ferromagnetic} (F). Contrary to the spin nematic phase, the F phase
breaks time reversal invariance but preserves parity $P^{(1)}_S \otimes P^{(2)}_L$.

Shifting the field $\Phi_-$ by $\sqrt{\pi K_-} / 4$ and passing to dimensionless
notations for the coupling constant,
\[
z_2 = 2 \sqrt{\pi K_+}\frac{\lambda_2 \alpha}{v}
\]
and total magnetization
\[
{\cal S}_+ = \frac{2 \pi \alpha}{\sqrt{K_+}} m^z,
\]
we write the variational energy density as
\be
{\cal E} = \frac{1}{2} \left( M^2 _- + {\cal S}_+ ^2 \right)
\mp z_2 {\cal S}_+ M^{1/K_-}_- .
\ee
Its minimization yields a finite gap in the (--) channel,
\be
M_- = K_- ^{- \frac{K_-}{2(K_- - 1)}} z^{\frac{K_-}{K_- - 1}}_2
\label{gap-F}
\ee
which supports a nonzero magnetization directed along the exchange anisotropy axis:
\be
{\cal S}_+ = \pm \sqrt{K_-} M_- = 
\pm K_- ^{- \frac{1}{2(K_- - 1)}} z^{\frac{K_-}{K_- - 1}}_2
\label{S_F}
\ee
The (+) channel remains gapless. Together with the finite spontaneous magnetization
this circumstance makes the longitudinal spin correlations algebraic {\sl and}
incommensurate:
\bea
\la S^z _{1} (x) S^z _{1, 2} (0) \ra &=& \la S^z \ra^2 -  \frac{K_+}{8\pi^2}
 \frac{1}{x^2}\nn\\
&+& {\rm const} \frac{(-1)^{x/a_0}}{|x|^{K_+/2}} \cos q_0 x,
\label{long-corr-F}
\eea
where $q_0 = {\cal S}_+ \sqrt{K_+}/2\alpha$. The transverse spin correlations
are short-ranged.

The ground state is doubly degenerate: the two vacua transforming to each other
under time reversal. The corresponding topological kinks have a finite mass gap
and carry the relative spin
\[
S^z _- = \frac{1}{\sqrt{\pi K_-}} \int_{-\infty}^{\infty} \rd x~\p_x \Phi_-
= \pm \frac{1}{2}.
\]

\begin{figure}[h]
\centering
\psfrag{K1}{$K_+$}\psfrag{K2}{$K_-$}
\psfrag{2}{2}\psfrag{1}{1}\psfrag{0}{0}\psfrag{a}{}\psfrag{b}{}
\psfrag{MSN}{MSN}\psfrag{CSN}{CSN}\psfrag{J}{J}\psfrag{Dimer}{D}
\psfrag{g}{$\Delta$}
\includegraphics[height=4cm]{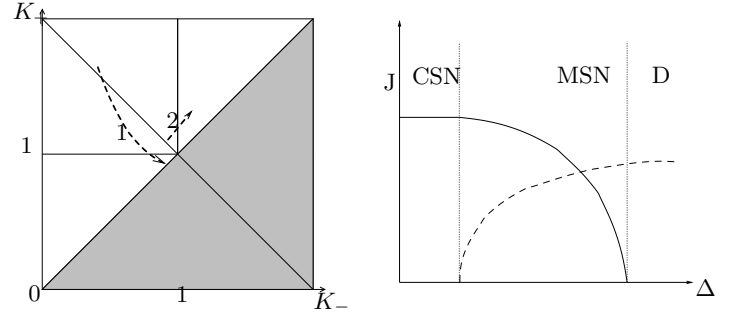}
\caption{The first path  shows one possible way of 
going through CSN, MSN and
D phases when $z_{\bot}<z_1$, by increasing
$\Delta$ and keeping $\Delta'$ to be a constant. The right
figure shows the qualitative change of the spin current (solid line)
and the demerization order parameter (dashed line) following this path. 
  In the second path
by decreasing $\Delta'$ and keeping  $\Delta$ at constant value,
the ladder enters the F phase.}
\end{figure}
\bigskip

\section{RG approach at  A-D boundary}

The results of sections IV and VI are valid far enough from the boundary
between sectors A and D where one of the two twist operators, ${\cal O}_{1,2}$, 
is strongly relevant while the other is strongly irrelevant.
On the boundary $K_- = 1$ separating these sectors both twist operators
become marginal. Therefore we can expect that 
in the immediate vicinity of the boundary,
\be
K_- = 1 - \delta_-, ~~~|\delta_-| \ll 1, \label{delta-}
\ee
far away from the SU(2)-symmetric point,  i.e.
$K_+ > 1$ in the sense that  $K_+ - 1 = O(1)$, 
the infrared behavior of model will be  
controlled by the 
interplay between the two parity-breaking operators
with a nonzero conformal spin, ${\cal O}_{1,2}$, 
and the longitudinal (conformal-scalar) terms 
$\p_x \Phi^{\pm}_R \p_x \Phi^{\pm}_L $ responsible for
renormalization of the coupling constants. 
So the starting low-energy
model should therefore contain both twist terms:
\bea
{\cal H} &=& {\cal H}^{(+)}_0 + {\cal H}^{(-)}_0 \nn\\
&+& \gamma_1 \p_x \Theta_+ \sin \tilde{\beta} \Theta_-
+  \gamma_2 \p_x \Phi_+ \sin \beta \Phi_- .
\label{new-model}
\eea
Here $\gamma_{1,2}$ differ from $\lambda_{1,2}$ by some multiplicative factors, 
and
\be
\tilde{\beta} = \frac{4\pi}{\beta} = \sqrt{4\pi K_-}
= \sqrt{4\pi} \left[ 1 + \frac{\delta_-}{2} + O(\delta^2 _-) \right]
\label{betas}
\ee

Notice that the two twist terms in (\ref{new-model}) contain vertex operators
(the sines) of mutually dual and {\sl nonlocal} fields, $\Theta_-$ and $\Phi_-$.
Models of this kind cannot be treated by the variational method used in the preceeding
sections. 
This is why in this section we address the RG flow of this model which will be studied
using a mapping
of the bosonic Hamiltonian (\ref{new-model}) onto a theory of four 
interacting real (Majorana) fermions
(c.f. Ref.\onlinecite{nge}).
\medskip

Let us first make all perturbations in (\ref{new-model}) strictly
marginal. This can be done by the following rescaling of the fields
in the (--) and (+) sectors:
\bea
\Phi_- \rightarrow \sqrt{K_-} \Phi_-, &&
\Theta_- \rightarrow (1/\sqrt{K_-} ) ~\Theta_- \label{rescaling-}\\
\Phi_+ \rightarrow \sqrt{q_+} \Phi_+, &&
\Theta_+ \rightarrow (1/\sqrt{q_+}) ~\Theta_+ \label{rescaling+}
\eea
The meaning of the first rescaling is transparent: we enforce the twist operators
in (\ref{new-model}) to have the scaling dimension 2.
This rescaling generates a current-current term in the (--) channel.
On the other hand,
fusing the two twist operators, one generates a similar term in the (+) sector;
that will renormalize the parameter $K_+$,
$
K_+ \rightarrow \bar{K}_+ = K_+ q_+ .
$
Below we will set
$
q_+ = 1 - \delta_+
$ 
assuming that $|\delta_+| \ll 1$. The Hamiltonian (\ref{new-model}) then acquires
the form:
\bea
{\cal H} &=& \frac{u}{2} \sum_{s=\pm} \left[\left(\p_x \Phi_s\right)^2
+ \left(\p_x \Theta_s\right)^2 \right]
- 2u \sum_{s=\pm} \delta_s \p_x \Phi_{sR}\p_x \Phi_{sL}\nn\\
&+& \gamma_1 \p_x \Theta_+ \sin \sqrt{4\pi} \Theta_-
+ \gamma_2 \p_x \Phi_+ \sin \sqrt{4\pi} \Phi_-
\label{back-rescaled-ham}
\eea
where $\delta_{\pm}$ satisfy the initial conditions
\be
\delta^{(0)}_+ = 0, ~~~\delta^{(0)}_- = \delta_-
\label{initial-cond} 
\ee
It is understood that extra factors appearing due to rescaling of $K_+$ are
absorbed into a redefinition of the coupling constants $\gamma_1$ and
$\gamma_2$.
\medskip

The structure of (\ref{back-rescaled-ham}) immediately suggests mapping onto
four real (Majorana) fermions, $\xi^a ~(a=0,1,2,3)$. 
This can be done using a correspondence
$$
\left( \Phi_+, \Theta_+ \right) \Rightarrow \left( \xi_1, \xi_2 \right),
~~~
\left( \Phi_-, \Theta_- \right) \Rightarrow \left( \xi_3, \xi_0 \right).
$$
and standard fermionization rules for the currents and vertex operators.
The resulting theory is given by the Euclidean action 
describing four degenerate massless fermions with a {\sl chirally asymmetric}
interaction:
\bea
&&S = 
\sum_{a=0}^3 \int \rd^2 z~
(\xi^a \bp \xi^a + \bxi^a \p \bxi^a) \nn\\
&& + 2\pi v \int \rd^2 z~
[ \delta_+~ \xi_1 \xi_2 \bxi_1 \bxi_2 
+ \delta_-~ \xi_3 \xi_0 \bxi_3 \bxi_0 \nn\\
&&~~~ +~  \gamma_+ (\xi_1 \xi_2 \xi_3 \bxi_0 + \bxi_1 \bxi_2 \bxi_3 \xi_0)\nn\\
&& ~~~+ ~ \gamma_- (\xi_1 \xi_2 \xi_0 \bxi_3 + \bxi_1 \bxi_2 \bxi_0 \xi_3)].
\label{maj-ham-int}
\eea
Here $\xi_a (z)$ and $\bxi_a (\bz)$ are holomorphic (left) and antiholomorphic (right)
components of the Majorana fields, 
$z = v\tau + \ri x$ and $\bz =v\tau - \ri x$ are complex coordinates,
$\p = \p/\p z$, $\bp = \p / \p \bz$,
and
\be
\gamma_{\pm} = \frac{\pi^{3/2} \alpha}{2\pi u} \left( \gamma_1 \pm 
\gamma_2 \right)
\ee

Due to its chiral asymmetry, the interaction in (\ref{maj-ham-int})
gives rise to renormalization of the velocities already on the one-loop level.
For this reason
we will discriminate between the velocities of different Majorana species,
and set
$
v_a = v ( 1 - 4\pi  \rho_a) ~(a=0,1,2,3),
$
where the dimensionless parameters
$\rho_a$ are subject to renormalization with
initial conditions
$\rho^{(0)}_a = 0.$

Using the standard fusion rules for fermion fields\cite{CFT}, one can easily derive
the following one-loop RG equations:
\bea
\dot{\delta_+} = - 2 \gamma_+ \gamma_- , &&
\dot{\delta_-} = 0,   \label{rg1}\\
\dot{\gamma_+} = \delta_- \gamma_- , &&
\dot{\gamma_-} = \delta_- \gamma_+ , \label{rg2}\\
\dot{\rho}_1 = \dot{\rho}_2 &=& \gamma^2_+  + \gamma^2 _- , \label{rg3}\\
\dot{\rho}_3 = \gamma^2 _+ ,&&
\dot{\rho}_0 = \gamma^2 _- ,\label{rg4}
\eea
where $\dot{g} \equiv \rd g(l)/\rd l$, $l = \ln (L/\alpha)$.

First of all, we observe that the coupling constant $\delta_-$ stays
unrenormalized:
\be
\delta_- (l) = \delta_- (0) = \delta_- \label{sol1}
\ee
Representing $\lambda_{\pm}$ as
$$
\gamma_{\pm} = g_1 \pm g_2 , ~~~
g_{1,2} = (\pi^{3/2} \alpha/2\pi u) \gamma_{1,2}
$$
we rewrite the first, third and fourth RG equations as
\bea
\dot{\delta_+} &=& - 2 \left( g^2 _1 - g^2 _2 \right) \label{RG1}\\
\dot{g}_1 = \delta_- g_1 , &&
\dot{g}_2 = -  \delta_- g_2 \label{RG2}
\eea
We see that, depending on the sign of $\delta_-$, either $g_1(l)$
or $g_2(l)$ grow up upon renormalization:
\medskip

\noindent
(a) \underline{$~~\delta_- > 0$}
\be
g_1 (l) = g^{(0)}_1 e^{\delta_- l}, ~~~
g_2 (l) = g^{(0)}_2 e^{- \delta_- l} \rightarrow 0
\label{regime-1}
\ee
Strong-coupling behavior of $g_1 (l)$ in (\ref{regime-1})
is associated with a dynamical  generation of a mass gap
\be
m_1 \propto |g_1|^{1/\delta_-}. \label{mass1}
\ee
\medskip

\noindent
(b) \underline{$~~\delta_- < 0$}
\be
g_2 (l) = g^{(0)}_2 e^{|\delta_-| l}, ~~~
g_1 (l) = g^{(0)}_1 e^{- |\delta_-| l} \rightarrow 0
\label{regime-2}
\ee
Here the mass gap is estimated as
\be
m_2 \propto |g_2|^{1/|\delta_-|}. \label{mass2}
\ee

The cases (a) and (b) describe the CSN and F phases,
respectively. 
Estimations (\ref{mass1}), (\ref{mass2}) are consistent with
the power-law scaling of the corresponding mass gaps, Eqs.(\ref{mass-sn}) and 
(\ref{gap-F}).
In both cases, $|\delta_+ (l)|$ flows to strong coupling. It goes to large
negative values in the case (a), implying that $K_+$ becomes even larger
upon renormalization. In the case (b) it flows to large positive values; so
the effective $K_+$ significantly
reduces, and that might indicate the importance of the neglected twist operator
${\cal O}_3$. Stability of the F phase is therefore under
question.
\medskip

Exactly at the boundary between sectors A and D
$\delta_- = 0$. In this case both $g_1$
and $g_1$ stay unrenormalized. Moreover,
\be
\delta_+ (l) = - 2 (g^2 _1 - g^2 _2) l.
\ee
So, if in addition we set $g_1 = g_2$, the 1-loop RG will display a 
weak-coupling regime for all coupling constants. This is the self-dual
point of the model where the interaction is not
renormalized:
for all effective couplings parametrizing interaction the $\beta$--function vanishes.
Amaizingly, in this case the Majorana action (\ref{maj-ham-int}) decouples into
two chirally asymmetric, independent parts, $S = S_{\rm I} + S_{\rm II}$, where
\be
S_{\rm I} = \int \rd^2 z~\left( \sum_{a=1}^3 \xi^a \bp \xi^a + \bxi^0 \p \bxi^0
+ g\xi^1 \xi^2 \xi^3 \bxi^0 \right)
\label{so3-action}
\ee
and $S_{\rm II}$ is obtained from $S_{\rm I}$ by reversing the chiralities of all
the fields. 
Notice that even though the present case corresponds to an essentially
anisotropic regime (remember that we are far away from the SU(2)-symmetric point
of the model), the effective theory on the boundary between
sectors A and D ($K_- = 1$) with the self-duality condition  $g_1 = g_2$ exhibits
an enlarged, chiral SO(3) $\otimes$ SO(3) symmetry.
Consistent with this symmetry is renormalization of the velocities.
The velocity of the singlet fermion, $\bxi^0$, stays intact: $\dot{\rho}_0 = 0$.
However, the triplet velocity is renormalized.
The RG equation 
$\dot{\rho}_t = 4 g^2$
$(\rho_i \equiv \rho_t$ $i=1,2,3$)
shows that $4\pi \rho_t (l)$ increases upon renormalization and reaches values
of the order of 1 in the region where $g^2 l \sim 1$. This sets up an infrared
energy scale in the problem, $\omega_0 \sim \Lambda \exp (- {\rm const}/g^2)$,
at which the triplet collective excitations soften significantly.
This is in agreement with the exact results for the spectrum
of model (\ref{so3-action}), recently obtained by Tsvelik \cite{tsvelik}.  

Interestingly enough, the 
exact solution \cite{tsvelik} shows that the chiral SO(3) symmetry of the action
(\ref{so3-action}) is spontaneously broken at T=0, and the
ground state of the model represents a ``chiral ferromagnet''
characterized by a nonzero  
expectation value of the vector current: 
\[
\la  {\bf I} \ra \neq 0, ~~~
I^a  = - (i/2) \eps^{abc} \xi_b \xi_c.
\]
Similarly, for action $S_{\rm II}$
\[
\la  \bar{\bf I} \ra \neq 0, ~~~
\bar{I}^a  = - (i/2) \eps^{abc} \bar{\xi}_b \bar{\xi}_c.
\]
 As long as the actions $S_{\rm I}$ and $S_{\rm II}$
are decoupled, there is no correlation between 
$\la  {\bf I} \ra$ and $\la  \bar{\bf I} \ra$, or equivalently, between
the magnetization 
$
{\bf m} =  \la  {\bf I} \ra + \la  \bar{\bf I} \ra
$
and spin current
$
{\bf j} = \la  {\bf I} \ra - \la  \bar{\bf I} \ra.
$
 Such correlation
appears upon deviation from the A--D boundary since in this case 
chirally-symmetric terms that couple  the actions $S_{\rm I}$ and $S_{\rm II}$
(and also introduce a finite XXZ anisotropy) are generated.
Thus, in the A-vicinity of the A--D boundary
$m_z \neq 0$, $j_z = 0$,
whereas in the D-vicinity the situation is just inverted:
$m_z = 0$, $j_z \neq 0$.
So, the resulting picture at the A--D boundary depends
on the side from which this boundary is approached,
implying that the CSN -- F transition is first-order. 
\medskip

Even though the action $S = S_{\rm I} +  S_{\rm II}$ provides the simplest field-theoretical
model for a frustrated ladder
with a chirally asymmetric interaction and, hence,
is quite interesting in its own right, we will refrain from its further discussion
because it does not account for the low-energy properties
of the zigzag spin-1/2 ladder with the {\sl generic} SU(2) symmetry 
(the point $K_+ = K_- = 1$).

\section{Conclusions}

In this paper we have analyzed the phase diagram of the spin-1/2 anisotropic zigzag ladder 
with a weak interchain coupling ($J'\ll J$).
Using the Abelian bosonization method combined with a variational approach, 
we have found that, depending on the anisotropy parameters, 
the system occurs either in the parity  and time-reversal symmetric,
spontaneously dimerized phase, or in one of 
those phases in which either parity is spontaneously broken while time reversal preserved,
or vice versa. These are the
critical and massive spin nematic phases
(CSN,MSN) and the critical ferromagnetic (F) phase. 
We have shown that the CSN phase extends well beyond
the XX limit and covers  broad regions A and B in the parameter space of the XXZ model
(see Figs.2). Each of these phases is characterized by topological excitations
carrying fractional quantum numbers.  

Starting from a vicinity of the XX point, we  addressed the 
nature of the transition between the CSN and D phases taking place
upon increasing the intra-chain anisotropy parameter $\Delta$, say, at a fixed positive
value of $\Delta'$. Typical curves are shown in Fig.2. In these two Figures we show
two possible scenarios whose realization depends on the 
ratio ($z_1 / z_{\perp}$) between the amplitudes of the main competing perturbations -- 
the twist operator ${\cal O}_1$ and the dimerization
field ${\cal D}$. 
The reason we considered each of these scenarios on equal footing
is due to the fact that the relationship between the parameters of our bosonized model, 
Eq.(\ref{cont-model-final}), and those of the microscopic Hamiltonian (\ref{ham})
is nonuniversal and, hence, known only by the order of magnitude. 
If $z_{\perp} > z_1$, the CSN -- D transition is first order. In the
opposite case, $z_{\perp} < z_1$,   
the CSN and D phases are sandwiched by the MSN phase characterized
by the coexistence of a finite spin current with dimerization. 
Then the variational approach uanmbiguously shows that CSN--MSN and MSN -- D transitions
are continuous, even though it is inadequate to identify their universality classes.
We believe that the final choice between the two possibilities discussed in this paper
will be made in future
numerical work (The accuracy of the recent DMRG calculations \cite{kabu,kabu1} 
for the S=1/2 zigzag ladders was reported
to be inadequte to resolve this issue. Moreover, 
the non-linear $\sigma$ model approach of  
Ref.~\onlinecite{kole},
which excludes the massive chiral phase for half integer spins, is only valid in
the vicinity of the classical Liftshitz point $j=1/4$.)

\medskip

Starting from the region A occupied by the CSN phase, one can also keep the in-chain anisotropy
intact and vary continuously the interchain anisotropy. In particular, one can smoothly go
from the case  of an antiferromagnetic
interchain coupling ($g_{\parallel} > 0$) to the case of a ferromagnetic coupling ($g_{\parallel} < 0$)
(see Fig.1).
We have shown that in such situation the ladder crosses over from the CSN phase to the F phase, 
the latter being dual to the former.

\acknowledgements
We acknowledge stimulating discussions on these and related topics with Fabian
Essler, Alexander Gogolin and Alexei Tsvelik. 
Two of us (M.F. and A.A.N.) are partly supported by MIUR,
under project
COFIN2003 "Field Theory, Statistical Mechanics and Electron Systems".

\appendix
\section{More about sector C}

In this Appendix we address the role of the so far neglected twist perturbation 
$\lambda_3 {\cal O}_3$  which becomes relevant in sector C. Adding this term to 
the effective
Hamiltonian leads us to the following theory:
\bea
{\cal H}_C &=& {\cal H}^{(+)}_0 + {\cal H}^{(-)}_0\nn\\
&-& \frac{\lambda_{\perp}}{\pi\alpha} \cos\sqrt{4\pi K_+} \Phi_+ \cos\sqrt{4\pi K_-} \Theta_-
\nn\\
&+& \frac{2\lambda_1}{\sqrt{K_+}} \p_x \Theta_+ \sin\sqrt{4\pi K_-} \Theta_- \nn\\
&+& \frac{\lambda_3}{\sqrt{K_-}} \p_x \Phi_- \sin \sqrt{4\pi K_+} \Phi_+ .
\label{ham-C}
\eea
It is convenient to introduce dimensionless notations for the coupling constant
\[
z_3 = 2 \sqrt{\frac{\pi}{K_-}} \frac{\lambda_3 \alpha}{v}
\]
and  the z-component of the relative spin density,
\[
{\cal Q}_- = 2 \sqrt{\pi} \alpha \p_x \Phi_-.
\]
The variational approach we followed in section V is straightforwardly generalized
for the present case. As compared to section V, here we have two additional variational
parameters: the relative spin density ${\cal Q}_-$ and a mixing angle $\zeta$
for the field $\Phi_+$. The variational energy then depends on six variables:
\bea
{\cal E}_C &=&\frac{1}{2} (M^2 _+ + M^2 _- + {\cal J}^2 _+ + {\cal Q}^2 _-) \nn\\
&\mp& z_1 {\cal J}_+ M^{K_-} _- \sin \gamma
\mp z_3 {\cal Q} M^{K_+} _+ \sin \zeta \nn\\
&-& z_{\perp}  M^{K_+} _+ M^{K_-} _- \cos \gamma \cos \zeta .
\label{var-energy-C}
\eea
Its minimization yields the following set of equations:
\bea
&&M^{K_-}_{-} (z_{\perp} M^{K_+}_{+} \sin \gamma \cos \zeta
\mp z_1 {\cal J}_+ \cos \gamma) = 0, \nn\\
&&M^{K_+}_{+} (z_{\perp} M^{K_-}_{-} \cos \gamma \sin \zeta
\mp z_3 {\cal Q}_- \cos \zeta) = 0, \nn\\
&&{\cal J}_+ \mp z_1 M^{K_-}_{-} \sin \gamma = 0, \nn\\
&&{\cal Q}_- \mp z_3 M^{K_+}_{+} \sin \zeta = 0, \nn\\
&&M_- ( 1 \mp z_1 {\cal J}_+ K_- M^{K_- - 2}_- \sin \gamma \nn\\
&& ~~~- z_{\perp} K_- M^{K_+}_{+}M^{K_- - 2}_- \cos \gamma \cos \zeta)
= 0, \nn\\
&&M_+ (1 \mp z_3 {\cal Q}_- K_+ M^{K_+ - 2}_+ \sin \zeta \nn\\
&& ~~~- z_{\perp} K_+ M^{K_-}_{+}M^{K_+ - 2}_+ \cos \gamma \cos \zeta)
= 0
\label{eqs-C}
\eea
 
There  exist solutions of these equations in which the second twist perturbation
$\lambda_3 {\cal O}_3$ plays no role:

(i) $\gamma = \zeta = 0~$, ${\cal J}_+ = {\cal Q}_- = 0~$ -- D phase;

(ii) $\gamma = \pi/2~$, $\zeta = 0~$, ${\cal Q}_- = 0~$ -- CSN phase;;

(iii) $0< \gamma < \pi/2~$, $\zeta = 0~$, ${\cal Q}_- = 0~$ -- MSN phase.
\medskip

\noindent
There exists a pair of solutions which are ``dual'' to (ii) and (iii), i.e.
can be obtained from the latter by the replacements
$z_1 \to z_3,$ $ {\cal J}_+ \to {\cal Q}_-,$ $ M_+ \leftrightarrow M_-$:

(iv)  $\gamma = 0~$, $\zeta = \pi/2~$, ${\cal J}_+ = 0$. This is a critical
phase with a nonzero relative magnetization (CRM), ${\cal Q}_- \neq 0$;

(v) $\gamma = 0~$, $0< \zeta < \pi/2~$, ${\cal J}_+ = 0~$ -- the massive
version of the above phase  (MRM).
In these two phases the twist operator ${\cal O}_1$ plays no role.

There also exist solutions in which both twist perturbations are effective. 
One of them corresponds to the case

(vi) $\gamma = \pi/2~$, $\zeta = \pi/2~$,
in which the $z_{\perp}$-perturbation is ineffective and the variational
energy decouples into a direct sum ${\cal E}_{CSN} + {\cal E}_{CRM}$.
The resulting phase is fully gapped and represents a mixture of CSN and CRM
phases -- mixed (M) phase with nonzero ${\cal J}_+$  and ${\cal Q}_-$.

The case of arbitrary values of the mixing angles, $\gamma, \zeta \neq 0,\pi/2$, should be
abandoned because, as follows from Eqs.(\ref{eqs-C}), it requires that
$z^2 _1 z^2 _3 = z^2 _{\perp}$, -- a condition which represents just a point
in the parameter space of the model and which, on the other hand, cannot be 
satisfied for all coupling constants being of the same order.

The minimal value of the variational energy is again given by Eq.(\ref{D.phase-energy}).
From this expression it is obvious that in sector C ($K_{\pm} < 1$) the M phase
has a lower energy than each of its ``constituents'', i.e. CSN and CRM phases. 
So we are left to find out if the M and MRM phases can compete with
the D and MSN phase.

Consider the MRM phase assuming the most favorable condition $z_3 > z_{\perp}$.
Since the  MRM phase is ``dual'' to the MSN phase, from Eq.(\ref{msn-range})
we can read off the range where it can exist:
\be
 1 -  (1-K_+) \frac{\ln (z_3/z_{\perp})}{\ln (1/z_3)} < K_- < 1.
\label{MRM-range}
\ee
We see that, except for an extremely unrealistic case $z^2 _3 > z_{\perp}$,
the condition (\ref{MRM-range}) determines a vicinity of the negative
semiaxis $K_- = 1$, $K_+ < 1$,  which is located in the unphysical part of
the $(K_+ , K_- )$ plane, well beyond sector C. Thus the MRM phase
should be abandoned.

Let us compare the energies of the M and MSN phases. In both cases the mass
$M_-$ is given by the same expression, so we only need to compare the masses
in the (+) channel. Comparing the mass $M_+$ in the M phase, $M_+ \sim z_3 ^{1/(1-K_+)}$,
with that in the MSN phase, Eq. (\ref{mass+:B}), we find that, except for
extremely small values of $z_{\perp}$, namely $z_{\perp} < z_1 z_3$, the
MSN phase is always more favorable.

Finally, we are left to compare the energies of the M and D phases.
On one hand, in the D phase we are below the line (\ref{msn-range}).
This means that $z_{\bot}^{1\over 2-k_D}>z_1^{1\over 1-K_-}$ implying that
the mass gap of the D phase is greater than the mass $M_-$ of the M phase.
On the other hand, to the left of the MSD-D transition line (\ref{MRM-range})
we have the condition  $z_{\bot}^{1\over 2-k_D}>z_3^{1\over 1-K_+}$
that tells us that the mass of the D phase is greater than the mass $M_+$ of the M phase.
Consequently, the D phase is energetically more favorable than the M phase.




\begin{thebibliography}{99}
\bibitem{com1} It should be remembered that
it is only at the SU(2) point that all three components
of the vector current satisfy the SU(2)$_1$ Kac-Moody algebra. In the XXZ
case ($\Delta < 1$) $J^x _{R,L}$ and $J^y _{R,L}$ acquire
anomalous dimensions, and only $J^z _{R,L}$ remain as the generators of
U(1)$_{R,L}$.
\end{thebibliography}
\end{document}